\documentclass[aps,prl,notitlepage,twocolumn]{revtex4-1}
\usepackage[dvips]{graphicx}% Include figure files
\usepackage{dcolumn}% Align table columns on decimal point
\usepackage{bm}% bold math
\usepackage{amsmath}
\usepackage[utf8]{inputenc}

\begin{document}

\title{Spin orbit coupling at the level of a single electron}

\author{V. F. Maisi}
\author{A. Hofmann}
\author{M. R\"{o}\"{o}sli}
\author{J. Basset}
\author{C. Reichl}
\author{W. Wegscheider}
\author{T. Ihn}
\author{K. Ensslin}
\affiliation{Solid State Physics Laboratory, ETH Zurich, CH-8093 Zurich, Switzerland}

\date{\today}

\begin{abstract}
We utilize electron counting techniques to distinguish a spin conserving fast tunneling process and a slower process involving spin flips in AlGaAs/GaAs-based double quantum dots. By studying the dependence of the rates on the interdot tunnel coupling of the two dots, we find that as many as $4 \%$ of the tunneling events occur with a spin flip related to spin-orbit coupling in GaAs. Our measurement has a fidelity of $99\ \%$ in terms of resolving whether a tunneling event occurred with a spin flip or not.
\end{abstract}

\maketitle

Spin-orbit coupling is an intrinsic property of any atom. In solids spin-orbit interaction may couples distant bands. It is of fundamental importance for the operation and performance of spin qubits~\cite{Ono2002,Johnson2005a,Petta2005,Koppens2006,Johnson2005,Fujita2015,Burkard2002} and for realizing Majorana fermions~\cite{Lutchyn2010,Mourik2012,Chang2015}. For instance, Stepanenko and coworkers proposed to use spin-orbit interaction for performing a CNOT operation of two spin qubits~\cite{Stepanenko2003}. In the experiments reported here we determine the strength of spin-orbit interaction in GaAs by measuring the spin dependent transfer of individual electrons between two tunnel coupled quantum dots, which are possible building blocks of qubits. We study electron tunneling between the dots in a regime where they are isolated from the electronic reservoirs. We measure the charge state of both dots simultaneously by employing real-time charge-sensing techniques~\cite{Vandersypen2004,Schleser2004,Kung2012}. This allows us to distinguish spin-flips taking place inside a quantum dot and spin-flips arising during tunneling processes between the two quantum dots. Our technique thus offers an efficient way of studying spin-flipping in quantum dots. We find that as many as $4 \%$ of the tunneling events involve a spin-flip mediated by spin-orbit interaction. This value, measured without driving the system externally, is an order of magnitude higher than that obtained previously for photon assisted processes~\cite{Braakman2014}.

\begin{figure}[t!]
	\centering

	\includegraphics[width=0.50\textwidth]{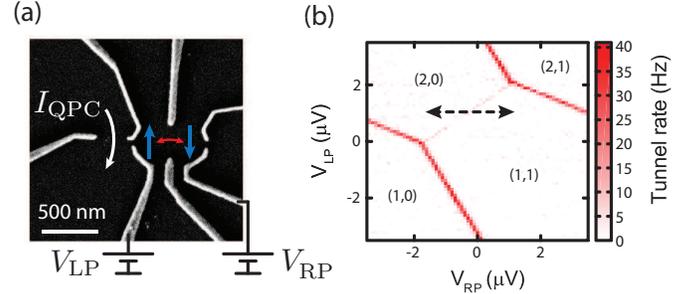}
	\caption{\label{fig:dev} (a), Scanning electron micrograph of the DQD. The light gray gates deplete the underlying 2DEG and form two quantum dots holding two electrons shown as blue arrows. The electrons may tunnel between the dots indicated by the red arrow. The tunneling events are detected via a nearby QPC by measuring the current $I_\mathrm{QPC}$ through it. (b), The rate of electron tunneling events around the $(2,0) \leftrightarrow (1,1)$ transition.}
\end{figure}

The double quantum dot (DQD) device is shown in Fig. \ref{fig:dev} (a). Metallic top gates (gray in the micrograph) are used for confining electrons into the two dots in a two dimensional electron gas lying $90\ \mathrm{nm}$ below the surface at a GaAs/AlGaAs interface. We tune both dots to the last electron~\cite{Tarucha1996,Ciorga2000}. Gate voltages $V_\mathrm{LP}$ and $V_\mathrm{RP}$ control the electron numbers in the left and right dot respectively. The other four gates surrounding the dots tune the tunneling rates between the two dots and between one of the dots and the source or drain reservoir below $100\ \mathrm{Hz}$. The additional gate on the left side of the DQD forms a quantum point contact (QPC). The current $I_\mathrm{QPC}$ transmitted through the QPC detects the charge state of each dot in real time. Measuring it allows us to determine the number of electrons in each dot and the tunneling rates separately for all processes~\cite{Vandersypen2004,Schleser2004,Kung2012}. 

Our study focuses on the region shown in Fig. \ref{fig:dev} (b), which presents the rate of tunneling as a function of $V_\mathrm{RP}$ and $V_\mathrm{LP}$. We observe four regions with stable electron numbers $(N_L,N_R)$ in the left and right dots (Ref~\citealp{vanderWiel2002}). At the boundaries of these regions (red lines), electron tunneling takes place. The bright red kinked lines near the top right and bottom left corners are caused by processes in which the total electron number in the double dot changes by tunneling to or from the reservoirs. Connecting the kinks of these two lines, we have a fainter straight line defining the boundary between the (2,0) and (1,1) states (charge degeneracy line). By restricting our measurements to the dashed black line cutting the charge degeneracy line, the tunneling processes involving the reservoirs are energetically forbidden due to the large Coulomb energy of the order of $1\ \mathrm{meV}$. Therefore the double dot is an isolated system restricted to the two occupation number combinations $(2,0)$ and $(1,1)$. In a similar way we measure the DQD at the transition $(1,0) \leftrightarrow (0,1)$.

\begin{figure}[t!]
	\centering
	\begin{tabular}{ll}
		\includegraphics[width=0.240\textwidth]{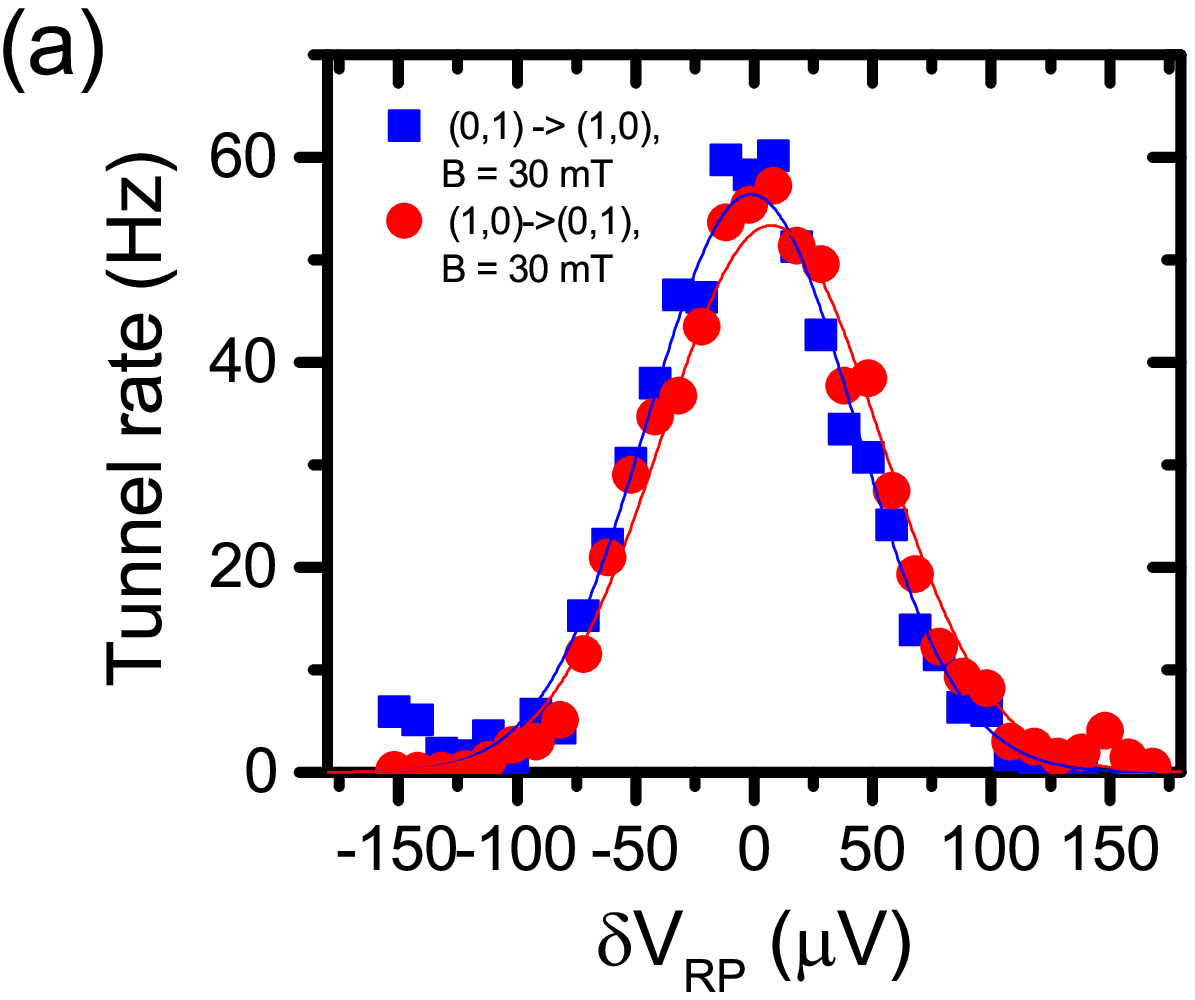} &
		\includegraphics[width=0.240\textwidth]{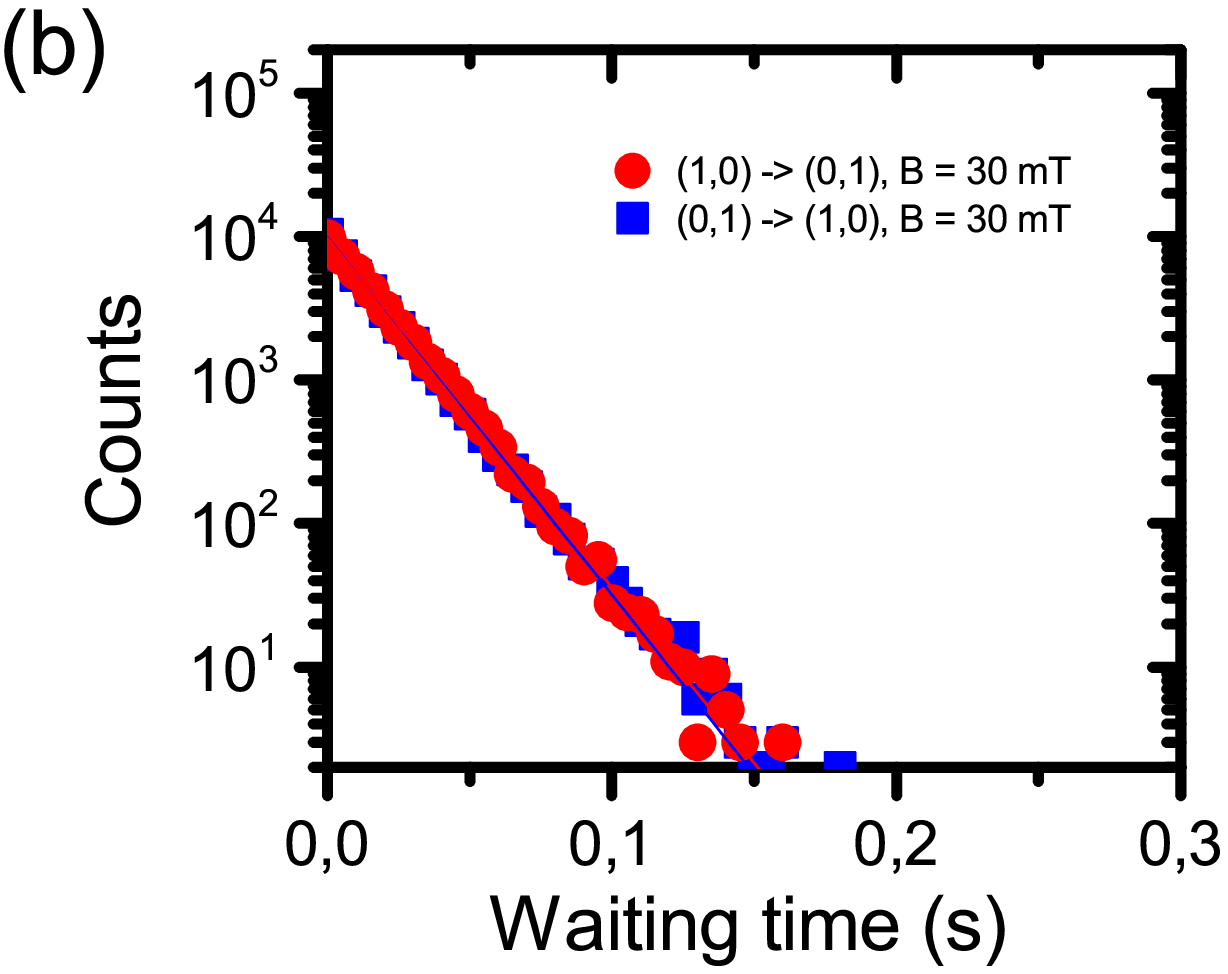} \\
		\includegraphics[width=0.240\textwidth]{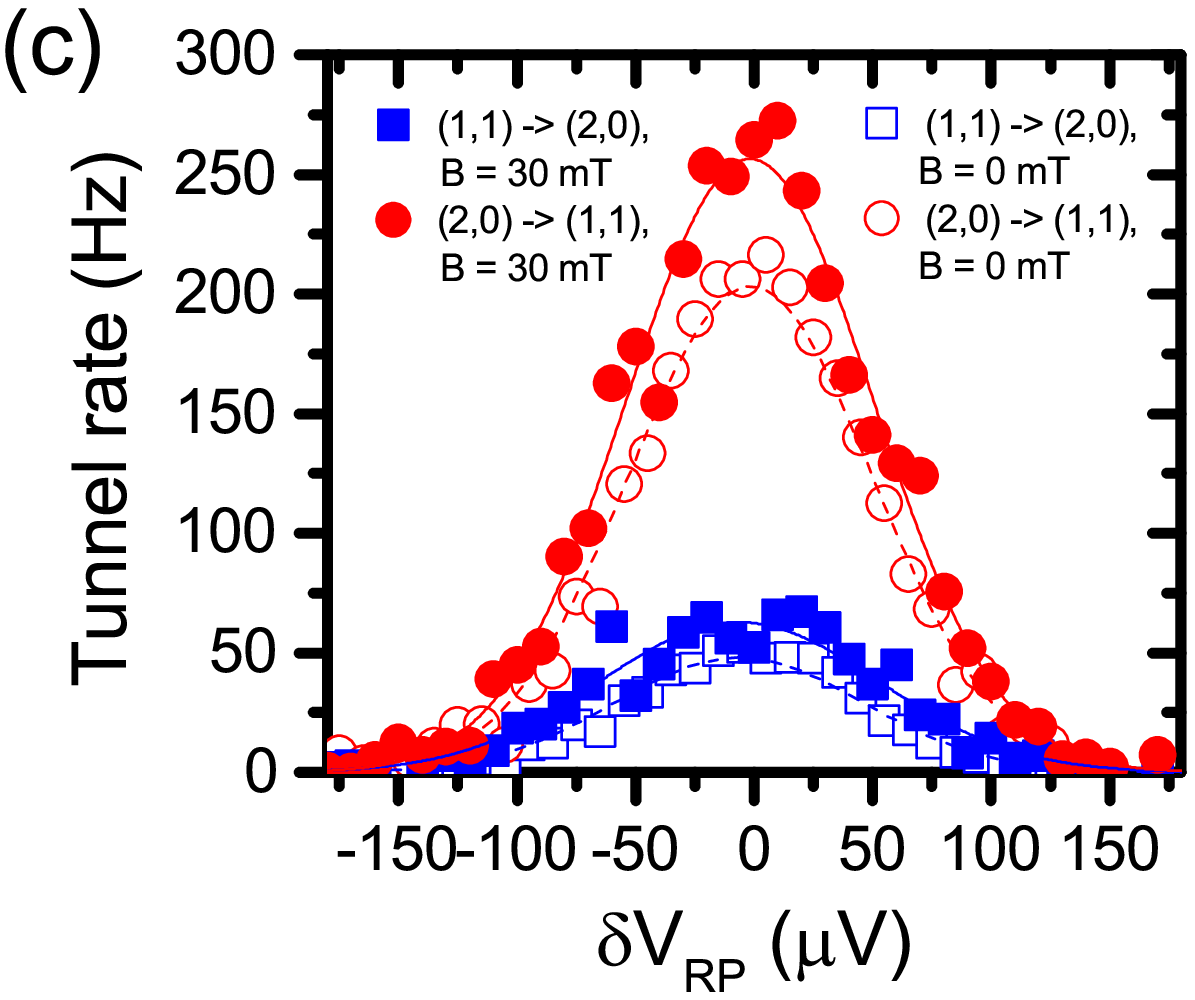} &
		\includegraphics[width=0.240\textwidth]{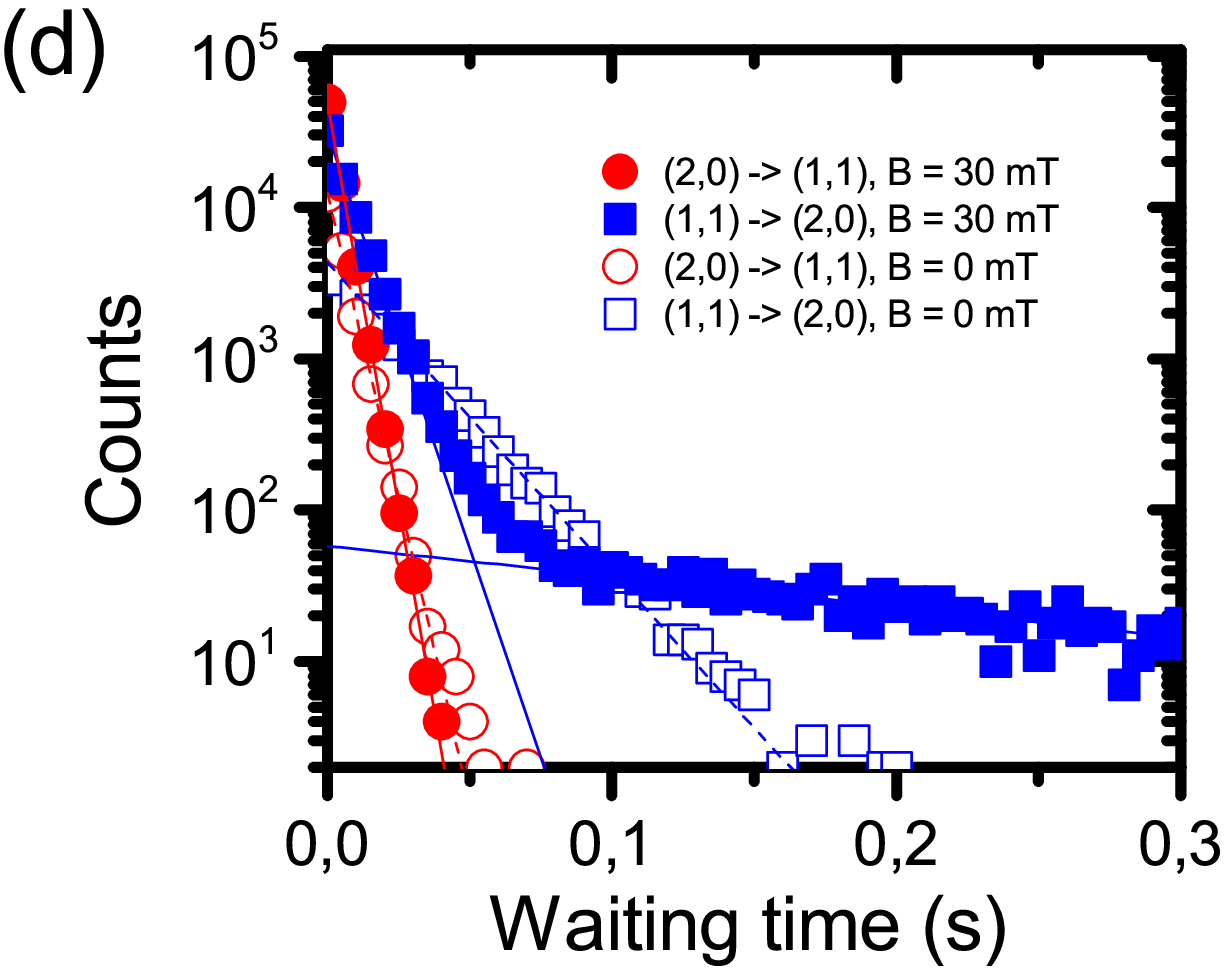} \\
	\end{tabular}
	\includegraphics[width=0.45\textwidth]{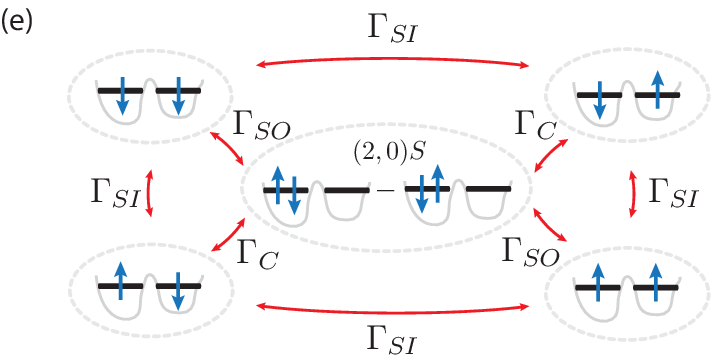}
	\begin{tabular}{ll}
		\includegraphics[width=0.3\textwidth]{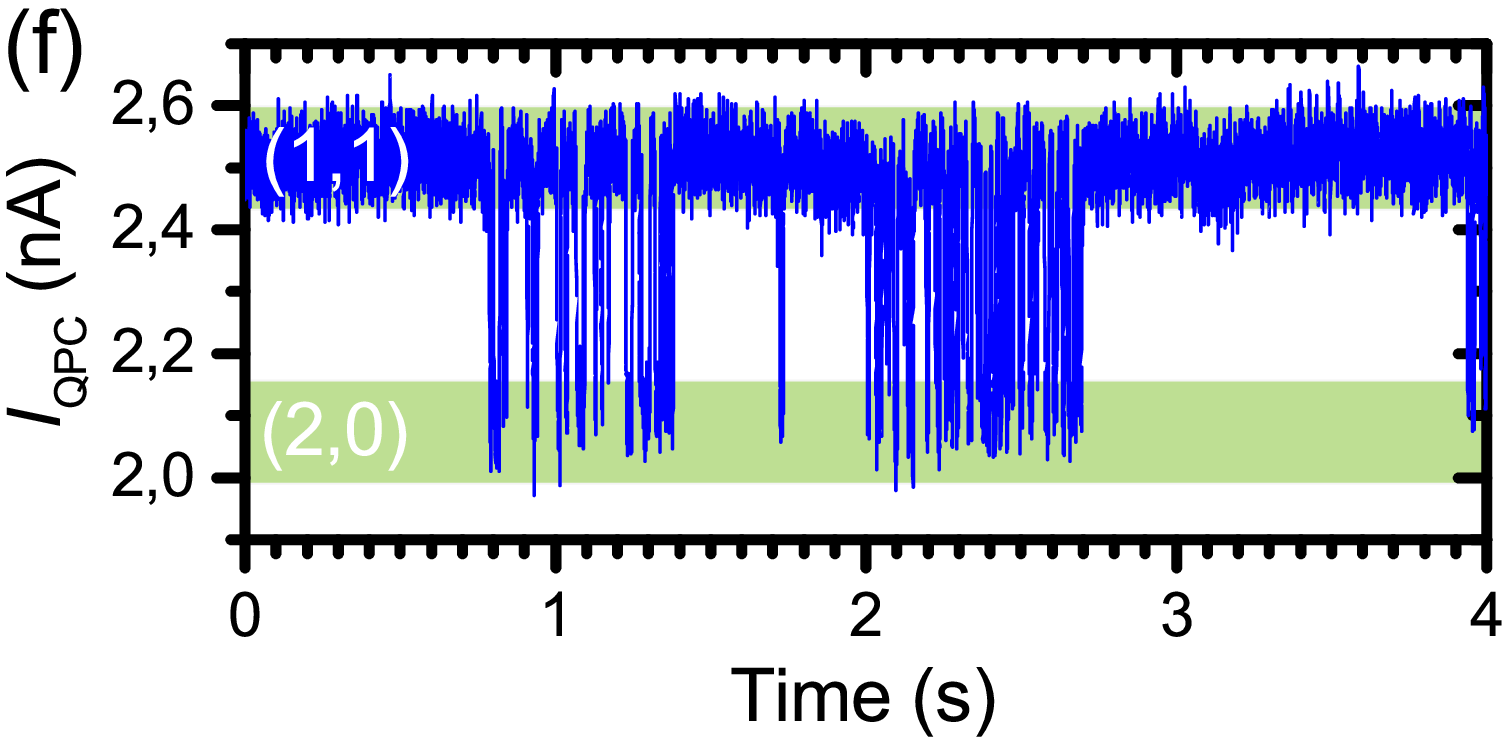} &
		\includegraphics[width=0.15\textwidth]{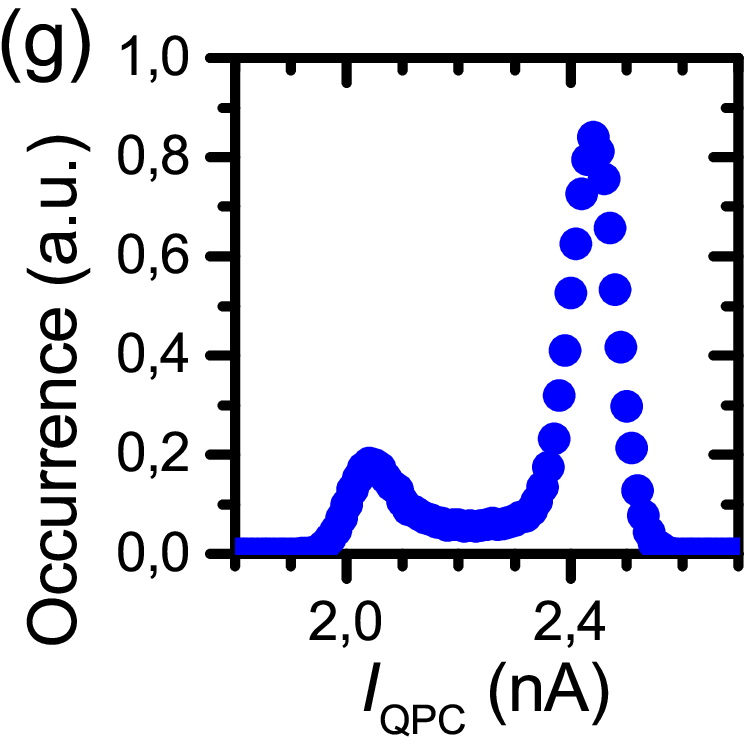}
	\end{tabular}
	\caption{\label{fig:idottunn} (a), Tunneling rates as a function of the energy detuning of the levels, changed with the plunger gate voltage $V_\mathrm{RP}$, for the transition $(1,0) \rightarrow (0,1)$ (red circles) and $(0,1) \rightarrow (1,0)$ (blue squares) with $B = 30\ \mathrm{mT}$. Solid lines show Gaussian fits with tunneling rates of $56 \pm 2\ \mathrm{Hz}$ and $53 \pm 2\ \mathrm{Hz}$ at the peak. (b), Waiting time distributions for the transitions of panel (a) with zero detuning. Solid lines are fits with average waiting times of $17.4 \pm 0.1\ \mathrm{ms}$ and $17.8 \pm 0.1\ \mathrm{ms}$. (c), Tunneling rates for transition $(2,0) \leftrightarrow (1,1)$ similarly as in panel (a). Solid symbols are with $B = 30\ \mathrm{mT}$ and open ones with $B = 0\ \mathrm{mT}$. Tunneling rates at the peak are $256 \pm 4\ \mathrm{Hz}$ and $63 \pm 2\ \mathrm{Hz}$, and $203 \pm 4\ \mathrm{Hz}$ and $48 \pm 2\ \mathrm{Hz}$ respectively. (d), Waiting time distributions of the transitions of panel (c) with zero detuning. The average waiting times obtained from the fits are $4.04 \pm 0.02\ \mathrm{ms}$, $7.89 \pm 0.04\ \mathrm{ms}$, $216 \pm 3\ \mathrm{ms}$, and $5.40 \pm 0.04\ \mathrm{ms}$ and $21.2 \pm 0.1\ \mathrm{ms}$ respectively.  (e), The five energetically allowed states around transition $(2,0) \leftrightarrow (1,1)$. The electrons move between the dots with a spin conserving process with a rate $\Gamma_C$ or a process involving a spin flip with a rate $\Gamma_{SO}$. The spins flip within the dots with a rate $\Gamma_{SI}$. (f), A time trace of the detector signal $I_\mathrm{QPC}$ showing bunching of the tunneling events. (g), Histogram of the time trace.}
\end{figure}

We first consider the transition $(1,0) \leftrightarrow (0,1)$. Figure \ref{fig:idottunn} (a) shows the tunneling rates in the two directions measured along a line cutting the charge degeneracy line similar to the dashed line in Fig. \ref{fig:dev} (b). The tunneling rates are determined from time traces by counting the number of tunneling events and dividing it by the total time spent in the corresponding initial state. We observe that the rates in the two directions are equal and that they form a resonance peak when $V_\mathrm{RP}$ is varied. Since the interdot tunnel coupling is very small, of the order of $ 1\ \mathrm{neV}$, the lineshape is defined by inelastic processes involving lattice vibrations and spurious electromagnetic fields. We find a gaussian line shape to fit the data well. Next we determine the waiting times of the $(1,0)$ and $(0,1)$ states again from the time traces. The waiting time statistics, measured at the position of the peak in Fig.~\ref{fig:idottunn} (a) are seen in Fig.~\ref{fig:idottunn} (b) to be distributed exponentially. The results are independent of applied magnetic field.

We perform the same analysis as before to the time traces measured at the transition $(2,0) \leftrightarrow (1,1)$. Figures~\ref{fig:idottunn} (c) and (d) present data for applied magnetic fields of $B = 30\ \mathrm{mT}$ (filled symbols) and $B = 0\ \mathrm{mT}$ (open symbols). In contrast to the last electron, the tunneling rates in the two directions are now not equal but differ by a factor of $4$. This factor arises from the degeneracies of the states. The $(2,0)$ ground state is a spin-singlet. The $(2,0)$ spin-triplet states are at energies much more than $kT$ above the singlet, because the single-particle excitation energy exceeds by far the exchange interaction. Thus, when both electrons reside in the same QD, they only occupy the spin singlet state $(2,0)S$. On the other hand, for the $(1,1)$ charge state, there are four spin states, which are degenerate because of the weak coupling of the QDs and the weak Overhauser field compared to temperature. At equilibrium, detailed balance leads to equipartitioning which makes sure that all four states have the same occupation probability irrespective of their tunneling coupling (Fig.~\ref{fig:idottunn} (g)). Therefore, the fourfold degeneracy of the $(1,1)$ charge state compared to the non-degenerate $(2,0)S$ state explains the factor of $4$ found between the tunneling rates in the two directions. A finite applied magnetic field of $30\ \mathrm{mT}$ leads to a $0.75\ \mathrm{\mu eV}$ Zeeman splitting of the states, which is still small compared to temperature ($5\ \mathrm{\mu eV}$)~\cite{LandoltBornstein2002}. Therefore, the ratio of $4$ in the tunneling rates persists.

We now inspect the waiting time distributions of the transition $(1,1) \rightarrow (2,0)$ presented in Fig. \ref{fig:idottunn} (d). We observe that the simple exponential distribution seen at $B = 0\ \mathrm{mT}$ (open squares) evolves into a distribution with two time scales at $B = 30\ \mathrm{mT}$ (filled squares). At zero applied magnetic field, the Overhauser field provides the quantization axis for the spins. Due to it's random fluctuations, the relative orientation of the two spins is random. A single exponential decay is obtained at zero magnetic field because the approximately 1 hour long measurement averages over random orientations of the Overhauser fields giving the same tunneling rate between the $(2,0)S$ and all $(1,1)$ states. With the 30 mT applied magnetic field, the external field dominates the Overhauser field and provides a common quantization axis for the spins in the two dots~\cite{Hanson2007}. The two time scales with the applied field arise from the fact that tunneling between the $(1,1)$ states and the $(2,0)S$ state sometimes requires a spin flip. The time scale for the spin flipping process is typically much longer than the spin conserving tunneling time, resulting in the so-called spin blockade~\cite{Ono2002,Johnson2005a,Petta2005,Koppens2006,Johnson2005,Fujita2015,Hanson2007,Fujita2015}. The two time scales are apparent in the typical time trace of the charge detector shown in Fig. \ref{fig:idottunn} (f) in which the spin conserving tunneling events are bunched into clear, distinct groups with long waiting times in-between such as from time $t = 2.7\ \mathrm{s}$ to $t = 3.9\ \mathrm{s}$.

Let us now assume that the fast tunneling rate is determined by transitions between states $(2,0)S$ and $(1,1)S$. In this case the tunneling rates are the same in forward and reverse direction, because both states are singly degenerate. The presence of three triplet states would not change the symmetry between forward and reverse processes significantly, because the spin-flip tunneling rate is very slow compared to the spin-conserving tunneling rate. Surprisingly, we observe the fast rate from $(1,1)$ to $(2,0)$ at 30 mT to be only half of the rate in the opposite direction (cf. the blue and red filled symbols in Fig. \ref{fig:idottunn} (d) and the corresponding fits), which refutes our assumption. We thus infer that the $(1,1)$ singlet and one of the $(1,1)$ triplet states mix fast on the time scale of spin-conserving tunneling. A convenient choice of the four $(1,1)$ states is the $\left| \uparrow \uparrow \right>$ ($T_+$), $\left| \uparrow \downarrow \right>$, $\left| \downarrow \uparrow \right>$, $\left| \downarrow \downarrow \right>$ ($T_-$) basis. The $\left| \uparrow \downarrow \right>$ and $\left| \downarrow \uparrow \right>$ states both tunnel couple to the $(2,0)S$ via a spin conserving process, which explains the ratio of two. This is in agreement with pioneering experiments of Petta and coworkers who found a characteristic spin mixing time scale of about $10\ \mathrm{ns}$ for $(1,1)S$ and $(1,1)T_0$ states, much smaller than the typical tunneling time in our experiment~\cite{Petta2005}. The spin blockade with the long waiting time occurs only when the system is in either the $\left| \uparrow \uparrow \right>$ or the $\left| \downarrow \downarrow \right>$ state which are subject to Zeeman splitting and thus can be effectively decoupled from the surrounding nuclear spins by suppressing the hyperfine interaction~\cite{Johnson2005,Koppens2005,Barthel2012}. 

Figure~\ref{fig:idottunn} (e) depicts the five quantum states that we have just introduced together with the spin-conserving tunneling rate $\Gamma_C$ and the spin-flip tunneling rate $\Gamma_{SO}$. In addition we consider spin-flips to take place not while tunneling but within a dot with a rate $\Gamma_{SI}$. This diagram was derived step by step from experimental results and seamlessly explains all integer ratios of tunneling rates at zero and finite magnetic field. In the zero field case the orientation of the spins follows the random Overhauser field of the two dots and is not strictly parallel or anti-parallel for the $(1,1)$ states. Therefore $\Gamma_{SO}$ and $\Gamma_C$ will average to the same number.

%Bunching occurs resulting in super-Poissonian statistic similar as other cases with two competing processes~\cite{Gustavsson2006a,Maisi2014}. For example the current noise is higher than for the case of a simple shot noise in the $(1,0) \leftrightarrow (0,1)$ transition.

\begin{figure}[t!]
	\centering
	\begin{tabular}{ll}
		\includegraphics[width=0.24\textwidth]{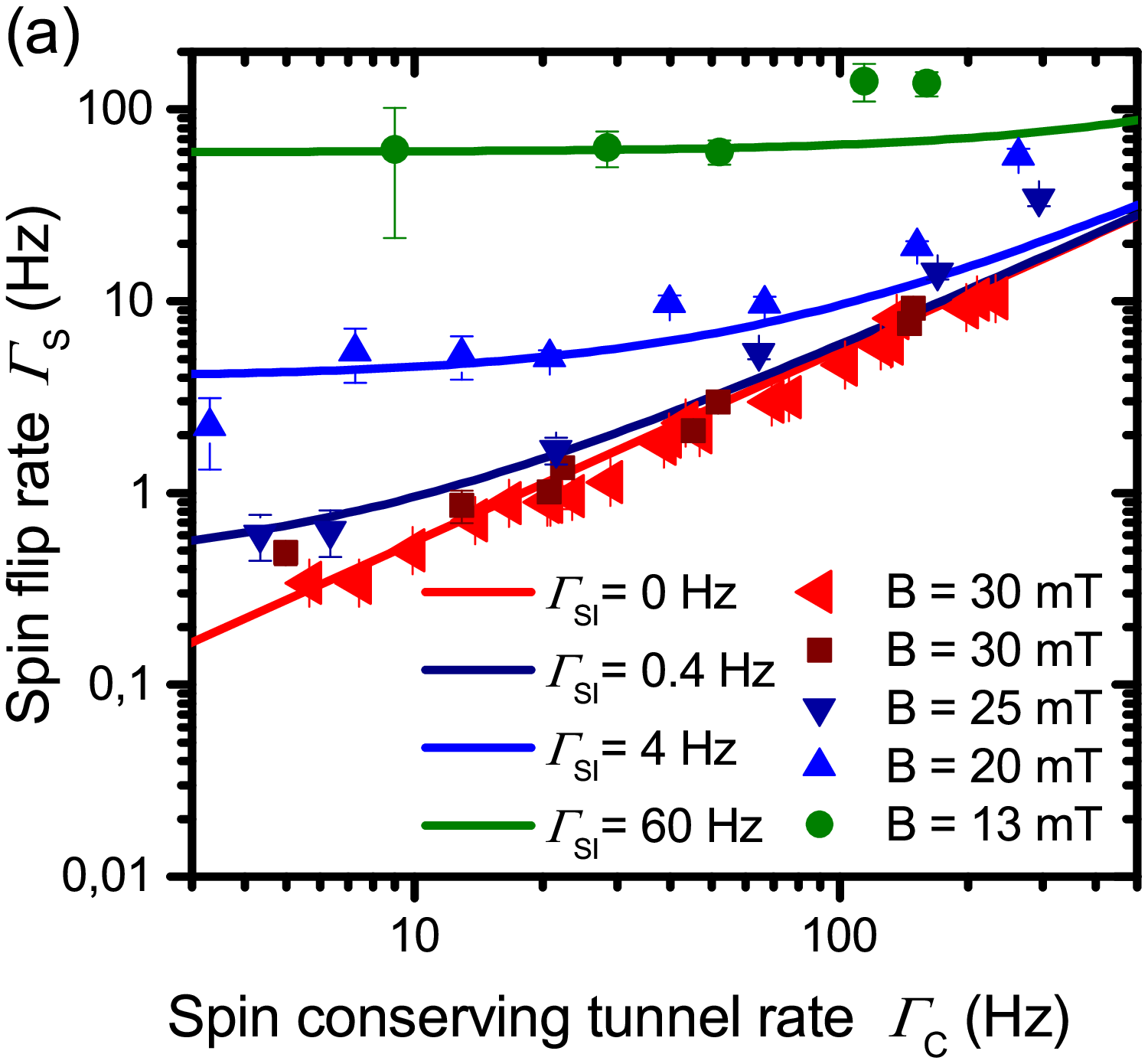}
		\includegraphics[width=0.24\textwidth]{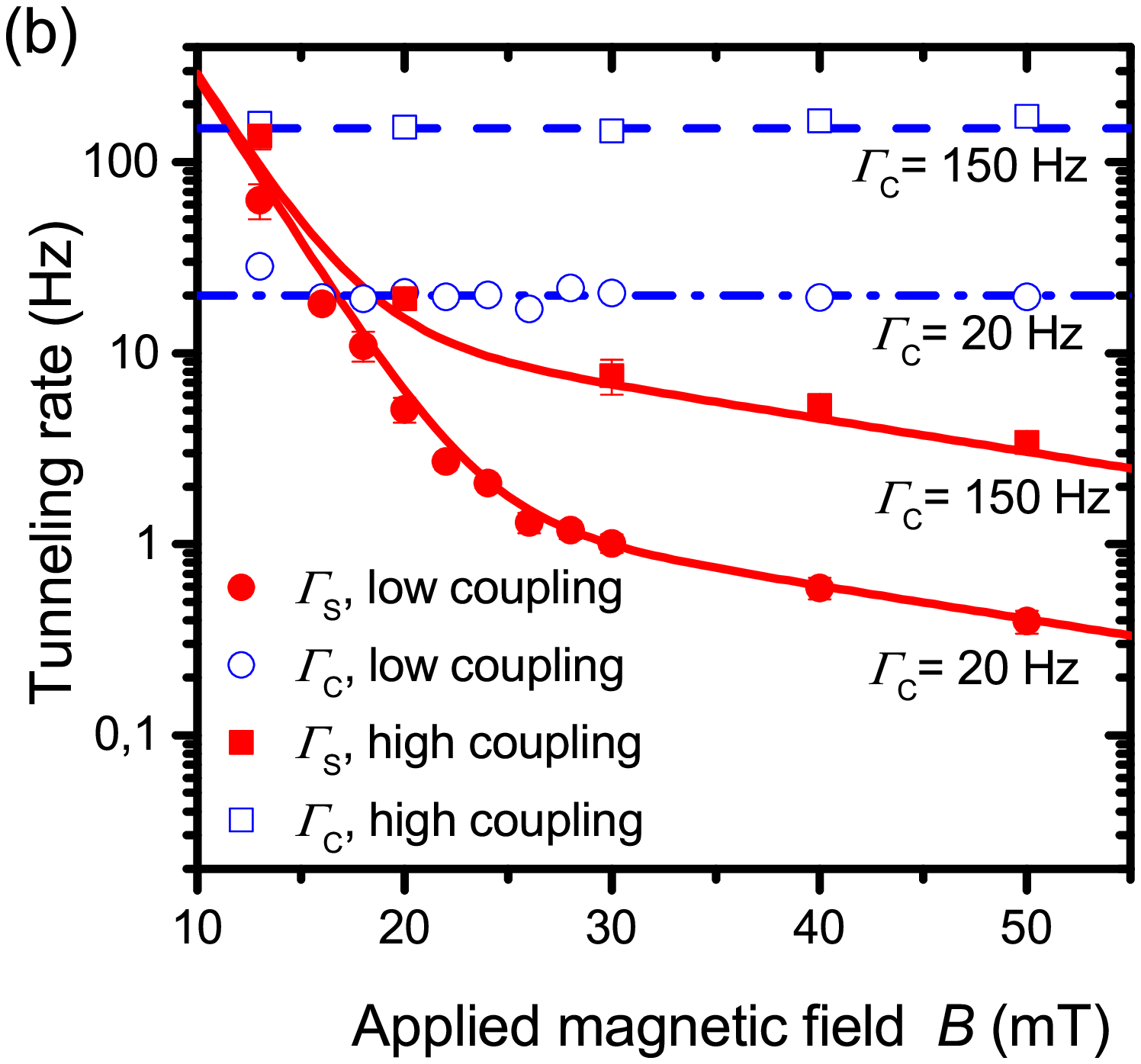}
	\end{tabular}
	\caption{\label{fig:tcoulpdep} (a), Spin flip rate $\Gamma_S$ as a function of the spin conserving tunneling rate $\Gamma_C$ at various applied fields $B$. The red straight line presents the linear dependence  $\Gamma_S = 0.04 \times \Gamma_C$ and the other lines are $\Gamma_S = 0.04 \times \Gamma_C + \Gamma_{SI}$ with $\Gamma_{SI}$ indicated in the figure. (b), Spin flip rate $\Gamma_S$ for transition $(1,1) \rightarrow (2,0)$ as a function of $B$ (filled red symbols) for two different tunnel couplings. The tunnel coupling is deduced from the tunneling rate $\Gamma_C$ of the $(2,0) \rightarrow (1,1)$ transition as presented with the open blue symbols. The solid red lines are guides for eyes with $\Gamma_S =15\ \mathrm{kHz} \cdot e^{-B/3\ \mathrm{mT}}+0.15\; \Gamma_C \cdot e^{-B/30\ \mathrm{mT}}$.}
\end{figure}

In Fig.~\ref{fig:tcoulpdep} (a) we present a measurement in which the tunnel coupling between the dots is changed with the barrier gate located between the two plunger gates. We measure the waiting times in the $(2,0) \leftrightarrow (1,1)$ transition in resonance and extract the fast spin-conserving tunneling rate $\Gamma_C$ and slow spin-flip rate from the two time-scales of the waiting time distributions (cf. Fig.~\ref{fig:idottunn} (e)).

The longer time-scale found in the experiment may originate from either the internal spin-flip rate within a dot ($\Gamma_{SI}$) or the spin-flip tunneling rate ($\Gamma_{SO}$).  In the case that the long time-scale is determined by $\Gamma_{SI}$, we need to substract the finite tunneling time ($1/\Gamma_C$) needed for detection of a spin-flip within a dot. The subtraction is not required if it is determined by $\Gamma_{SO}$. However, the subtraction introduces only a small error since $\Gamma_{SO} \ll \Gamma_C$. The modified spin-flip rate is called $\Gamma_S$ and may describe either process. We observe that $\Gamma_S$ is directly proportional to $\Gamma_C$ over almost two orders of magnitude at the highest field value. This dependence implies that the spin flip process involves electron tunneling and thus we have $\Gamma_S = \Gamma_{SO}$. When considering the theoretical prediction for a tunneling process invoking spin-orbit interaction with tunneling, the spin flip rate is expected to be $\Gamma_{SO} = (d/l_\mathrm{so})^2 \Gamma_C/2$, where $d$ is the distance between the dots and $l_\mathrm{so}$ the spin-orbit length~\cite{Khaetskii2000,Stepanenko2012,Danon2013}. Since in our case we have $\Gamma_{SO}/\Gamma_C = 0.04$, and the distance between the dots is approximately $l \sim 300\ \mathrm{nm}$, we estimate the spin-orbit length $l_\mathrm{so} = 1\ \mathrm{\mu m}$, which is consistent with expectations~\cite{Danon2013}. We rule out the hyperfine interaction to be dominant for spin-flip tunneling because it requires strong hybridization of the electronic levels of the two dots~\cite{Stepanenko2012}, which is absent in our weakly coupled system. Thus we demonstrate here the detection of the spin-orbit coupling at a level of single electrons. With the low magnetic fields of Fig.~\ref{fig:tcoulpdep} (a) we observe that the spin flip rate is not dependent on the tunnel coupling and saturates to $\Gamma_S = 2\Gamma_{SI}$ because of two possibilities of a spin-flip processes within a dot, see Fig.~\ref{fig:idottunn} (e). Our method allows for distinguishing between the spin flipping occuring predominantly during a tunneling event or by relaxation within a quantum dot. In Fig.~\ref{fig:tcoulpdep} (b) we present the magnetic field dependence of $\Gamma_S$ for two different tunnel couplings $\Gamma_C$. We observe a crossover from hyperfine interaction with no dependence on tunnel coupling but exponential dependence of $B$ at low fields to spin-orbit mediated spin-flipping with linear dependence on tunnel coupling but only a weak dependence on the applied field. The spin flip rates reported in Ref.~\citealp{Fujita2015} are consistent with our findings and the spin-flips take place with a tunneling process.

Let us now consider the fidelity of our spin-flip detection. We define a threshold time $\tau$. If the waiting time of the $(1,1)$ state (see Fig.~\ref{fig:idottunn} (d)) is longer that $\tau$, we deduce that a spin-flip occurred. For shorter waiting times, we infer that a spin conserving tunneling event occured. The cumulative distribution function of the exponential distribution yields the error probability of interpreting a spin-flip process as spin conserving one as $1-e^{-\Gamma_{S}\tau}$, which is the fraction of spin-flip events with waiting time smaller than $\tau$. Similarly the error of interpreting a spin conserving process as a spin-flip one is  $e^{-\Gamma_{C}\tau}$. Thus we obtain the fraction of events which are misinterpreted as $(1-p) e^{-\Gamma_C\tau} + p (1-e^{-\Gamma_{S}\tau})$, where $p = \Gamma_{S}/(\Gamma_C+\Gamma_{S})$, is the probability for a spin-flip process. For $\Gamma_{S} \ll \Gamma_C$, the error is minimized by $\tau = -\Gamma_C^{-1} \ln (p^2)$. For our experiment with $p = 0.04$, we thus detect $99\ \%$ of the tunneling events correctly by simply considering whether the waiting time was longer or shorter than the threshold $\tau$.

\begin{figure}[t!]
	\centering
	\includegraphics[width=0.50\textwidth]{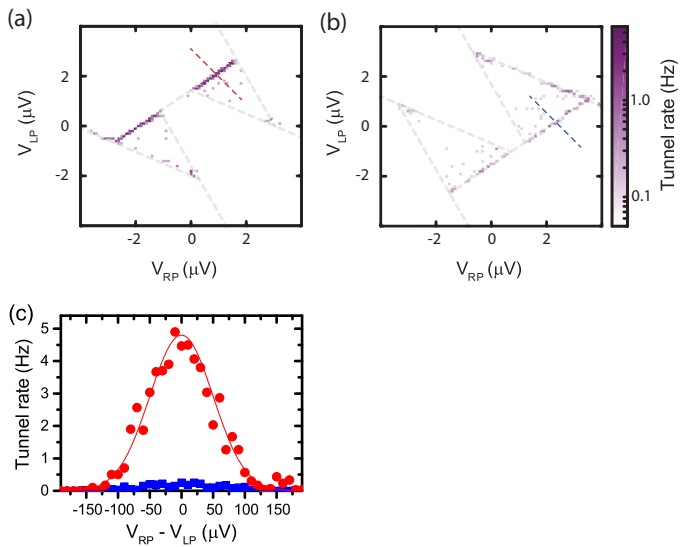}
	\caption{\label{fig:finbias} Current through the double dot as function of the plunger gate voltages $V_\mathrm{RP}$ and $V_\mathrm{LP}$ in panel (a) in the non-blocked direction, $V_b = +200\ \mathrm{\mu V}$, and spin blocked direction $V_b = +200\ \mathrm{\mu V}$, in panel (b). (c), Cuts along the lines indicated in (a) and (b). Solid lines show fits with rates at the peak of  $4.2 \pm 0.6\ \mathrm{Hz}$ and $0.2 \pm 0.1\ \mathrm{Hz}$}
\end{figure}

Finally, we illustrate the connection of our experiments to transport measurements performed previously~\cite{Ono2002,Johnson2005a,Koppens2006} and demonstrate that the spin-orbit interaction is probed by investigating the forward and reverse current of the double dot. We apply a bias voltage $V_b$ between the source and the drain contacts. This leads to a triangular region around the triple points of Fig.~\ref{fig:dev} (b) where current can flow between source and drain as shown in Figs.~\ref{fig:finbias} (a) and (b) for the two bias directions. We tune the interdot tunneling rate, $\Gamma_C \lesssim 10\ \mathrm{Hz}$, to be approximately an order of magnitude slower than the tunneling to source and drain in the allowed direction: $\Gamma_R \sim \Gamma_L \sim 60\ \mathrm{Hz}$. We use $B = 30\ \mathrm{mT}$ so that spin-orbit interaction is the dominant spin-flip mechanism. Since the interdot process is the slowest one, it limits the current. Current flows only if the energy levels in the two dots are resonant and inside the bias triangles. This gives rise to the two dark parallel lines of Fig.~\ref{fig:finbias} (a). For the reverse direction in panel (b), the lines are fainter because spin blockade suppresses transport. Next, we measure the rate at which electrons tunnel through the device across the dashed detuning lines shown in Fig.~\ref{fig:finbias} (a) and (b). The result is presented in panel (c) for forward direction (red) and backward direction (blue). Since $\Gamma_C \ll \Gamma_{L,R}$, these rates directly probe the rate at which electrons tunnel between the dots. These are the same rates as presented in Fig.~\ref{fig:idottunn} (c). The difference between the two cases is that at finite bias, the electrons in the dot are exchanged due to allowed tunneling to the reservoirs, so the system is open. Without the applied bias we have a closed system. We observe that the electron exchange has a striking impact on the results. Whereas the isolated case leads to ratio of $1:4$ for the tunneling rates, with the open system we obtain $1:20$. The difference is understood by considering the blocked direction. In the open system, if the spins are anti-parallel, electrons pass through the device quickly until a parallel spin enters from the reservoir. This leads to fast transition to the blocked state. In contrast, in the isolated case the blocked state is reached only by an electron spin flip. %Hence not only the lifting of the spin blockade as reported in Ref.~\citealp{Johnson2005a} but the blockade itself in view of suppressed current is a phenomenon requiring electron exchange which we demonstrate here experimentally. 
The current through the DQD in the two bias directions can be described using a standard rate equation with the states and rates as indicated in Fig.~\ref{fig:idottunn} (e) together with the $(2,1)$ state and the tunneling rates $\Gamma_L$ and $\Gamma_R$ to source and drain. With the assumptions $\Gamma_{SO} \ll \Gamma_C \ll \Gamma_R, \Gamma_L$, we obtain the forward current $I_\mathrm{fwd} = e\Gamma_C$ and the reverse current in the spin blocked direction of $I_\mathrm{rev} = 4e\Gamma_{SO}/3$. Thus from the ratio of the currents we obtain directly the ratio $\Gamma_{SO}/\Gamma_C = 3 I_\mathrm{rev}/4I_\mathrm{fwd}$. Averaging at the peak of Fig.~\ref{fig:finbias} (c) yields $I_\mathrm{fwd}/e = 4.8 \pm 0.1\ \mathrm{Hz}$ and $I_\mathrm{rev}/e = 0.24 \pm 0.01\ \mathrm{Hz}$. Thus we obtain $\Gamma_{SO}/\Gamma_C = 0.038 \pm 0.002$, which is consistent with our findings for the isolated case and prooves that the spin-orbit coupling is probed in the transport measurements provided that the dots are in the right regime of tunnel coupling.

In conclusion we have studied spin-orbit interaction at the level of a single-electron. We detected the spin-orbit interaction mediated tunneling with $99\ \%$ fidelity. The tunneling rates in the isolated double dot are insensitive to spin flipping and the nontrivial statistics of tunneling times needs to be analysed. For a transport measurement we demonstrated that the spin-orbit coupling is probed from the tunneling currents. Our technique is a new tool to concisely study spin-flip processes in quantum dots with the capability to distinguish spin-flip tunneling processes and the spin-flips within a quantum dot.

{\it Acknowledgements} We thank S. Tarucha and D. Loss for fruitful discussions.

\bibliographystyle{aipnum4-1}
\bibliography{referencesurl}

\begin{thebibliography}{25}%
\makeatletter
\providecommand \@ifxundefined [1]{%
 \@ifx{#1\undefined}
}%
\providecommand \@ifnum [1]{%
 \ifnum #1\expandafter \@firstoftwo
 \else \expandafter \@secondoftwo
 \fi
}%
\providecommand \@ifx [1]{%
 \ifx #1\expandafter \@firstoftwo
 \else \expandafter \@secondoftwo
 \fi
}%
\providecommand \natexlab [1]{#1}%
\providecommand \enquote  [1]{``#1''}%
\providecommand \bibnamefont  [1]{#1}%
\providecommand \bibfnamefont [1]{#1}%
\providecommand \citenamefont [1]{#1}%
\providecommand \href@noop [0]{\@secondoftwo}%
\providecommand \href [0]{\begingroup \@sanitize@url \@href}%
\providecommand \@href[1]{\@@startlink{#1}\@@href}%
\providecommand \@@href[1]{\endgroup#1\@@endlink}%
\providecommand \@sanitize@url [0]{\catcode `\\12\catcode `\$12\catcode
  `\&12\catcode `\#12\catcode `\^12\catcode `\_12\catcode `\%12\relax}%
\providecommand \@@startlink[1]{}%
\providecommand \@@endlink[0]{}%
\providecommand \url  [0]{\begingroup\@sanitize@url \@url }%
\providecommand \@url [1]{\endgroup\@href {#1}{\urlprefix }}%
\providecommand \urlprefix  [0]{URL }%
\providecommand \Eprint [0]{\href }%
\providecommand \doibase [0]{http://dx.doi.org/}%
\providecommand \selectlanguage [0]{\@gobble}%
\providecommand \bibinfo  [0]{\@secondoftwo}%
\providecommand \bibfield  [0]{\@secondoftwo}%
\providecommand \translation [1]{[#1]}%
\providecommand \BibitemOpen [0]{}%
\providecommand \bibitemStop [0]{}%
\providecommand \bibitemNoStop [0]{.\EOS\space}%
\providecommand \EOS [0]{\spacefactor3000\relax}%
\providecommand \BibitemShut  [1]{\csname bibitem#1\endcsname}%
\let\auto@bib@innerbib\@empty
%</preamble>
\bibitem [{\citenamefont {Ono}\ \emph {et~al.}(2002)\citenamefont {Ono},
  \citenamefont {Austing}, \citenamefont {Tokura},\ and\ \citenamefont
  {Tarucha}}]{Ono2002}%
  \BibitemOpen
  \bibfield  {author} {\bibinfo {author} {\bibfnamefont {K.}~\bibnamefont
  {Ono}}, \bibinfo {author} {\bibfnamefont {D.~G.}\ \bibnamefont {Austing}},
  \bibinfo {author} {\bibfnamefont {Y.}~\bibnamefont {Tokura}}, \ and\ \bibinfo
  {author} {\bibfnamefont {S.}~\bibnamefont {Tarucha}},\ }\href@noop {}
  {\bibfield  {journal} {\bibinfo  {journal} {Science}\ }\textbf {\bibinfo
  {volume} {297}},\ \bibinfo {pages} {1313} (\bibinfo {year}
  {2002})}\BibitemShut {NoStop}%
\bibitem [{\citenamefont {Johnson}\ \emph
  {et~al.}(2005{\natexlab{a}})\citenamefont {Johnson}, \citenamefont {Petta},
  \citenamefont {Marcus}, \citenamefont {Hanson},\ and\ \citenamefont
  {Gossard}}]{Johnson2005a}%
  \BibitemOpen
  \bibfield  {author} {\bibinfo {author} {\bibfnamefont {A.~C.}\ \bibnamefont
  {Johnson}}, \bibinfo {author} {\bibfnamefont {J.~R.}\ \bibnamefont {Petta}},
  \bibinfo {author} {\bibfnamefont {C.~M.}\ \bibnamefont {Marcus}}, \bibinfo
  {author} {\bibfnamefont {M.~P.}\ \bibnamefont {Hanson}}, \ and\ \bibinfo
  {author} {\bibfnamefont {A.~C.}\ \bibnamefont {Gossard}},\ }\href@noop {}
  {\bibfield  {journal} {\bibinfo  {journal} {Phys. Rev. B}\ }\textbf {\bibinfo
  {volume} {72}},\ \bibinfo {pages} {165308} (\bibinfo {year}
  {2005}{\natexlab{a}})}\BibitemShut {NoStop}%
\bibitem [{\citenamefont {Petta}\ \emph {et~al.}(2005)\citenamefont {Petta},
  \citenamefont {Johnson}, \citenamefont {Taylor}, \citenamefont {Laird},
  \citenamefont {Yacoby}, \citenamefont {Lukin}, \citenamefont {Marcus},
  \citenamefont {Hanson},\ and\ \citenamefont {Gossard}}]{Petta2005}%
  \BibitemOpen
  \bibfield  {author} {\bibinfo {author} {\bibfnamefont {J.~R.}\ \bibnamefont
  {Petta}}, \bibinfo {author} {\bibfnamefont {A.~C.}\ \bibnamefont {Johnson}},
  \bibinfo {author} {\bibfnamefont {J.~M.}\ \bibnamefont {Taylor}}, \bibinfo
  {author} {\bibfnamefont {E.~A.}\ \bibnamefont {Laird}}, \bibinfo {author}
  {\bibfnamefont {A.}~\bibnamefont {Yacoby}}, \bibinfo {author} {\bibfnamefont
  {M.~D.}\ \bibnamefont {Lukin}}, \bibinfo {author} {\bibfnamefont {C.~M.}\
  \bibnamefont {Marcus}}, \bibinfo {author} {\bibfnamefont {M.~P.}\
  \bibnamefont {Hanson}}, \ and\ \bibinfo {author} {\bibfnamefont {A.~C.}\
  \bibnamefont {Gossard}},\ }\href@noop {} {\bibfield  {journal} {\bibinfo
  {journal} {Science}\ }\textbf {\bibinfo {volume} {309}},\ \bibinfo {pages}
  {2180} (\bibinfo {year} {2005})}\BibitemShut {NoStop}%
\bibitem [{\citenamefont {Koppens}\ \emph {et~al.}(2006)\citenamefont
  {Koppens}, \citenamefont {Buizert}, \citenamefont {Tielrooij}, \citenamefont
  {Vink}, \citenamefont {Nowack}, \citenamefont {Meunier}, \citenamefont
  {Kouwenhoven},\ and\ \citenamefont {Vandersypen}}]{Koppens2006}%
  \BibitemOpen
  \bibfield  {author} {\bibinfo {author} {\bibfnamefont {F.~H.~L.}\
  \bibnamefont {Koppens}}, \bibinfo {author} {\bibfnamefont {C.}~\bibnamefont
  {Buizert}}, \bibinfo {author} {\bibfnamefont {K.-J.}\ \bibnamefont
  {Tielrooij}}, \bibinfo {author} {\bibfnamefont {I.~T.}\ \bibnamefont {Vink}},
  \bibinfo {author} {\bibfnamefont {K.~C.}\ \bibnamefont {Nowack}}, \bibinfo
  {author} {\bibfnamefont {L.~P.}\ \bibnamefont {Meunier}}, \bibinfo {author}
  {\bibfnamefont {L.~P.}\ \bibnamefont {Kouwenhoven}}, \ and\ \bibinfo {author}
  {\bibfnamefont {L.~M.~K.}\ \bibnamefont {Vandersypen}},\ }\href@noop {}
  {\bibfield  {journal} {\bibinfo  {journal} {Nature}\ }\textbf {\bibinfo
  {volume} {442}},\ \bibinfo {pages} {766} (\bibinfo {year}
  {2006})}\BibitemShut {NoStop}%
\bibitem [{\citenamefont {Johnson}\ \emph
  {et~al.}(2005{\natexlab{b}})\citenamefont {Johnson}, \citenamefont {Petta},
  \citenamefont {Taylor}, \citenamefont {Yacoby}, \citenamefont {Lukin},
  \citenamefont {Marcus}, \citenamefont {Hanson},\ and\ \citenamefont
  {Gossard}}]{Johnson2005}%
  \BibitemOpen
  \bibfield  {author} {\bibinfo {author} {\bibfnamefont {A.~C.}\ \bibnamefont
  {Johnson}}, \bibinfo {author} {\bibfnamefont {J.~R.}\ \bibnamefont {Petta}},
  \bibinfo {author} {\bibfnamefont {J.~M.}\ \bibnamefont {Taylor}}, \bibinfo
  {author} {\bibfnamefont {A.}~\bibnamefont {Yacoby}}, \bibinfo {author}
  {\bibfnamefont {M.~D.}\ \bibnamefont {Lukin}}, \bibinfo {author}
  {\bibfnamefont {C.~M.}\ \bibnamefont {Marcus}}, \bibinfo {author}
  {\bibfnamefont {M.~P.}\ \bibnamefont {Hanson}}, \ and\ \bibinfo {author}
  {\bibfnamefont {A.~C.}\ \bibnamefont {Gossard}},\ }\href@noop {} {\bibfield
  {journal} {\bibinfo  {journal} {Nature}\ }\textbf {\bibinfo {volume} {435}},\
  \bibinfo {pages} {925} (\bibinfo {year} {2005}{\natexlab{b}})}\BibitemShut
  {NoStop}%
\bibitem [{\citenamefont {Fujita}\ \emph {et~al.}(2015)\citenamefont {Fujita},
  \citenamefont {Morimoto}, \citenamefont {Kiyama}, \citenamefont {Allison},
  \citenamefont {Larsson}, \citenamefont {Ludwig}, \citenamefont {Valentin},
  \citenamefont {Wieck}, \citenamefont {Oiwa},\ and\ \citenamefont
  {Tarucha}}]{Fujita2015}%
  \BibitemOpen
  \bibfield  {author} {\bibinfo {author} {\bibfnamefont {T.}~\bibnamefont
  {Fujita}}, \bibinfo {author} {\bibfnamefont {K.}~\bibnamefont {Morimoto}},
  \bibinfo {author} {\bibfnamefont {H.}~\bibnamefont {Kiyama}}, \bibinfo
  {author} {\bibfnamefont {G.}~\bibnamefont {Allison}}, \bibinfo {author}
  {\bibfnamefont {M.}~\bibnamefont {Larsson}}, \bibinfo {author} {\bibfnamefont
  {A.}~\bibnamefont {Ludwig}}, \bibinfo {author} {\bibfnamefont {S.~R.}\
  \bibnamefont {Valentin}}, \bibinfo {author} {\bibfnamefont {A.~D.}\
  \bibnamefont {Wieck}}, \bibinfo {author} {\bibfnamefont {A.}~\bibnamefont
  {Oiwa}}, \ and\ \bibinfo {author} {\bibfnamefont {S.}~\bibnamefont
  {Tarucha}},\ }\href@noop {} {\bibfield  {journal} {\bibinfo  {journal}
  {Arxiv}\ ,\ \bibinfo {pages} {1504.03696}} (\bibinfo {year}
  {2015})}\BibitemShut {NoStop}%
\bibitem [{\citenamefont {Burkard}\ and\ \citenamefont
  {Loss}(2002)}]{Burkard2002}%
  \BibitemOpen
  \bibfield  {author} {\bibinfo {author} {\bibfnamefont {G.}~\bibnamefont
  {Burkard}}\ and\ \bibinfo {author} {\bibfnamefont {D.}~\bibnamefont {Loss}},\
  }\href {\doibase 10.1103/PhysRevLett.88.047903} {\bibfield  {journal}
  {\bibinfo  {journal} {Phys. Rev. Lett.}\ }\textbf {\bibinfo {volume} {88}},\
  \bibinfo {pages} {047903} (\bibinfo {year} {2002})}\BibitemShut {NoStop}%
\bibitem [{\citenamefont {Lutchyn}, \citenamefont {Sau},\ and\ \citenamefont
  {Das~Sarma}(2010)}]{Lutchyn2010}%
  \BibitemOpen
  \bibfield  {author} {\bibinfo {author} {\bibfnamefont {R.~M.}\ \bibnamefont
  {Lutchyn}}, \bibinfo {author} {\bibfnamefont {J.~D.}\ \bibnamefont {Sau}}, \
  and\ \bibinfo {author} {\bibfnamefont {S.}~\bibnamefont {Das~Sarma}},\ }\href
  {\doibase 10.1103/PhysRevLett.105.077001} {\bibfield  {journal} {\bibinfo
  {journal} {Phys. Rev. Lett.}\ }\textbf {\bibinfo {volume} {105}},\ \bibinfo
  {pages} {077001} (\bibinfo {year} {2010})}\BibitemShut {NoStop}%
\bibitem [{\citenamefont {Mourik}\ \emph {et~al.}(2012)\citenamefont {Mourik},
  \citenamefont {Zuo}, \citenamefont {Frolov}, \citenamefont {Plissard},
  \citenamefont {Bakkers},\ and\ \citenamefont {Kouwenhoven}}]{Mourik2012}%
  \BibitemOpen
  \bibfield  {author} {\bibinfo {author} {\bibfnamefont {V.}~\bibnamefont
  {Mourik}}, \bibinfo {author} {\bibfnamefont {K.}~\bibnamefont {Zuo}},
  \bibinfo {author} {\bibfnamefont {S.~M.}\ \bibnamefont {Frolov}}, \bibinfo
  {author} {\bibfnamefont {S.~R.}\ \bibnamefont {Plissard}}, \bibinfo {author}
  {\bibfnamefont {E.~P. A.~M.}\ \bibnamefont {Bakkers}}, \ and\ \bibinfo
  {author} {\bibfnamefont {L.~P.}\ \bibnamefont {Kouwenhoven}},\ }\href@noop {}
  {\bibfield  {journal} {\bibinfo  {journal} {Science}\ }\textbf {\bibinfo
  {volume} {336}},\ \bibinfo {pages} {1003} (\bibinfo {year}
  {2012})}\BibitemShut {NoStop}%
\bibitem [{\citenamefont {Chang}\ \emph {et~al.}(2015)\citenamefont {Chang},
  \citenamefont {Albrecht}, \citenamefont {Jespersen}, \citenamefont
  {Kuemmeth}, \citenamefont {Krogstrup}, \citenamefont {Nygard},\ and\
  \citenamefont {Marcus}}]{Chang2015}%
  \BibitemOpen
  \bibfield  {author} {\bibinfo {author} {\bibfnamefont {W.}~\bibnamefont
  {Chang}}, \bibinfo {author} {\bibfnamefont {S.~M.}\ \bibnamefont {Albrecht}},
  \bibinfo {author} {\bibfnamefont {T.~S.}\ \bibnamefont {Jespersen}}, \bibinfo
  {author} {\bibfnamefont {F.}~\bibnamefont {Kuemmeth}}, \bibinfo {author}
  {\bibfnamefont {P.}~\bibnamefont {Krogstrup}}, \bibinfo {author}
  {\bibfnamefont {J.}~\bibnamefont {Nygard}}, \ and\ \bibinfo {author}
  {\bibfnamefont {C.~M.}\ \bibnamefont {Marcus}},\ }\href@noop {} {\bibfield
  {journal} {\bibinfo  {journal} {Nature Nanotech.}\ }\textbf {\bibinfo
  {volume} {10}},\ \bibinfo {pages} {232} (\bibinfo {year} {2015})}\BibitemShut
  {NoStop}%
\bibitem [{\citenamefont {Stepanenko}\ \emph {et~al.}(2003)\citenamefont
  {Stepanenko}, \citenamefont {Bonesteel}, \citenamefont {DiVincenzo},
  \citenamefont {Burkard},\ and\ \citenamefont {Loss}}]{Stepanenko2003}%
  \BibitemOpen
  \bibfield  {author} {\bibinfo {author} {\bibfnamefont {D.}~\bibnamefont
  {Stepanenko}}, \bibinfo {author} {\bibfnamefont {N.~E.}\ \bibnamefont
  {Bonesteel}}, \bibinfo {author} {\bibfnamefont {D.~P.}\ \bibnamefont
  {DiVincenzo}}, \bibinfo {author} {\bibfnamefont {G.}~\bibnamefont {Burkard}},
  \ and\ \bibinfo {author} {\bibfnamefont {D.}~\bibnamefont {Loss}},\ }\href
  {\doibase 10.1103/PhysRevB.68.115306} {\bibfield  {journal} {\bibinfo
  {journal} {Phys. Rev. B}\ }\textbf {\bibinfo {volume} {68}},\ \bibinfo
  {pages} {115306} (\bibinfo {year} {2003})}\BibitemShut {NoStop}%
\bibitem [{\citenamefont {Vandersypen}\ \emph {et~al.}(2004)\citenamefont
  {Vandersypen}, \citenamefont {Elzerman}, \citenamefont {Schouten},
  \citenamefont {{Willems van Beveren}}, \citenamefont {Hanson},\ and\
  \citenamefont {Kouwenhoven}}]{Vandersypen2004}%
  \BibitemOpen
  \bibfield  {author} {\bibinfo {author} {\bibfnamefont {L.~M.~K.}\
  \bibnamefont {Vandersypen}}, \bibinfo {author} {\bibfnamefont {J.~M.}\
  \bibnamefont {Elzerman}}, \bibinfo {author} {\bibfnamefont {R.~N.}\
  \bibnamefont {Schouten}}, \bibinfo {author} {\bibfnamefont {L.~H.}\
  \bibnamefont {{Willems van Beveren}}}, \bibinfo {author} {\bibfnamefont
  {R.}~\bibnamefont {Hanson}}, \ and\ \bibinfo {author} {\bibfnamefont {L.~P.}\
  \bibnamefont {Kouwenhoven}},\ }\href@noop {} {\bibfield  {journal} {\bibinfo
  {journal} {Applied Physics Letters}\ }\textbf {\bibinfo {volume} {85}},\
  \bibinfo {pages} {4394} (\bibinfo {year} {2004})}\BibitemShut {NoStop}%
\bibitem [{\citenamefont {Schleser}\ \emph {et~al.}(2004)\citenamefont
  {Schleser}, \citenamefont {Ruh}, \citenamefont {Ihn}, \citenamefont
  {Ensslin}, \citenamefont {Driscoll},\ and\ \citenamefont
  {Gossard}}]{Schleser2004}%
  \BibitemOpen
  \bibfield  {author} {\bibinfo {author} {\bibfnamefont {R.}~\bibnamefont
  {Schleser}}, \bibinfo {author} {\bibfnamefont {E.}~\bibnamefont {Ruh}},
  \bibinfo {author} {\bibfnamefont {T.}~\bibnamefont {Ihn}}, \bibinfo {author}
  {\bibfnamefont {K.}~\bibnamefont {Ensslin}}, \bibinfo {author} {\bibfnamefont
  {D.~C.}\ \bibnamefont {Driscoll}}, \ and\ \bibinfo {author} {\bibfnamefont
  {A.~C.}\ \bibnamefont {Gossard}},\ }\href@noop {} {\bibfield  {journal}
  {\bibinfo  {journal} {Applied Physics Letters}\ }\textbf {\bibinfo {volume}
  {85}} (\bibinfo {year} {2004})}\BibitemShut {NoStop}%
\bibitem [{\citenamefont {K\"ung}\ \emph {et~al.}(2012)\citenamefont {K\"ung},
  \citenamefont {R\"ossler}, \citenamefont {Beck}, \citenamefont {Marthaler},
  \citenamefont {Golubev}, \citenamefont {Utsumi}, \citenamefont {Ihn},\ and\
  \citenamefont {Ensslin}}]{Kung2012}%
  \BibitemOpen
  \bibfield  {author} {\bibinfo {author} {\bibfnamefont {B.}~\bibnamefont
  {K\"ung}}, \bibinfo {author} {\bibfnamefont {C.}~\bibnamefont {R\"ossler}},
  \bibinfo {author} {\bibfnamefont {M.}~\bibnamefont {Beck}}, \bibinfo {author}
  {\bibfnamefont {M.}~\bibnamefont {Marthaler}}, \bibinfo {author}
  {\bibfnamefont {D.~S.}\ \bibnamefont {Golubev}}, \bibinfo {author}
  {\bibfnamefont {Y.}~\bibnamefont {Utsumi}}, \bibinfo {author} {\bibfnamefont
  {T.}~\bibnamefont {Ihn}}, \ and\ \bibinfo {author} {\bibfnamefont
  {K.}~\bibnamefont {Ensslin}},\ }\href@noop {} {\bibfield  {journal} {\bibinfo
   {journal} {Phys. Rev. X}\ }\textbf {\bibinfo {volume} {2}},\ \bibinfo
  {pages} {011001} (\bibinfo {year} {2012})}\BibitemShut {NoStop}%
\bibitem [{\citenamefont {Braakman}\ \emph {et~al.}(2014)\citenamefont
  {Braakman}, \citenamefont {Danon}, \citenamefont {Schreiber}, \citenamefont
  {Wegscheider},\ and\ \citenamefont {Vandersypen}}]{Braakman2014}%
  \BibitemOpen
  \bibfield  {author} {\bibinfo {author} {\bibfnamefont {F.~R.}\ \bibnamefont
  {Braakman}}, \bibinfo {author} {\bibfnamefont {J.}~\bibnamefont {Danon}},
  \bibinfo {author} {\bibfnamefont {L.~R.}\ \bibnamefont {Schreiber}}, \bibinfo
  {author} {\bibfnamefont {W.}~\bibnamefont {Wegscheider}}, \ and\ \bibinfo
  {author} {\bibfnamefont {L.~M.~K.}\ \bibnamefont {Vandersypen}},\ }\href
  {\doibase 10.1103/PhysRevB.89.075417} {\bibfield  {journal} {\bibinfo
  {journal} {Phys. Rev. B}\ }\textbf {\bibinfo {volume} {89}},\ \bibinfo
  {pages} {075417} (\bibinfo {year} {2014})}\BibitemShut {NoStop}%
\bibitem [{\citenamefont {Tarucha}\ \emph {et~al.}(1996)\citenamefont
  {Tarucha}, \citenamefont {Austing}, \citenamefont {Honda}, \citenamefont
  {van~der Hage},\ and\ \citenamefont {Kouwenhoven}}]{Tarucha1996}%
  \BibitemOpen
  \bibfield  {author} {\bibinfo {author} {\bibfnamefont {S.}~\bibnamefont
  {Tarucha}}, \bibinfo {author} {\bibfnamefont {D.~G.}\ \bibnamefont
  {Austing}}, \bibinfo {author} {\bibfnamefont {T.}~\bibnamefont {Honda}},
  \bibinfo {author} {\bibfnamefont {R.~J.}\ \bibnamefont {van~der Hage}}, \
  and\ \bibinfo {author} {\bibfnamefont {L.~P.}\ \bibnamefont {Kouwenhoven}},\
  }\href@noop {} {\bibfield  {journal} {\bibinfo  {journal} {Phys. Rev. Lett.}\
  }\textbf {\bibinfo {volume} {77}},\ \bibinfo {pages} {3613} (\bibinfo {year}
  {1996})}\BibitemShut {NoStop}%
\bibitem [{\citenamefont {Ciorga}\ \emph {et~al.}(2000)\citenamefont {Ciorga},
  \citenamefont {Sachrajda}, \citenamefont {Hawrylak}, \citenamefont {Gould},
  \citenamefont {Zawadzki}, \citenamefont {Jullian}, \citenamefont {Feng},\
  and\ \citenamefont {Wasilewski}}]{Ciorga2000}%
  \BibitemOpen
  \bibfield  {author} {\bibinfo {author} {\bibfnamefont {M.}~\bibnamefont
  {Ciorga}}, \bibinfo {author} {\bibfnamefont {A.~S.}\ \bibnamefont
  {Sachrajda}}, \bibinfo {author} {\bibfnamefont {P.}~\bibnamefont {Hawrylak}},
  \bibinfo {author} {\bibfnamefont {C.}~\bibnamefont {Gould}}, \bibinfo
  {author} {\bibfnamefont {P.}~\bibnamefont {Zawadzki}}, \bibinfo {author}
  {\bibfnamefont {S.}~\bibnamefont {Jullian}}, \bibinfo {author} {\bibfnamefont
  {Y.}~\bibnamefont {Feng}}, \ and\ \bibinfo {author} {\bibfnamefont
  {Z.}~\bibnamefont {Wasilewski}},\ }\href {\doibase
  10.1103/PhysRevB.61.R16315} {\bibfield  {journal} {\bibinfo  {journal} {Phys.
  Rev. B}\ }\textbf {\bibinfo {volume} {61}},\ \bibinfo {pages} {R16315}
  (\bibinfo {year} {2000})}\BibitemShut {NoStop}%
\bibitem [{\citenamefont {van~der Wiel}\ \emph {et~al.}(2002)\citenamefont
  {van~der Wiel}, \citenamefont {De~Franceschi}, \citenamefont {Elzerman},
  \citenamefont {Fujisawa}, \citenamefont {Tarucha},\ and\ \citenamefont
  {Kouwenhoven}}]{vanderWiel2002}%
  \BibitemOpen
  \bibfield  {author} {\bibinfo {author} {\bibfnamefont {W.~G.}\ \bibnamefont
  {van~der Wiel}}, \bibinfo {author} {\bibfnamefont {S.}~\bibnamefont
  {De~Franceschi}}, \bibinfo {author} {\bibfnamefont {J.~M.}\ \bibnamefont
  {Elzerman}}, \bibinfo {author} {\bibfnamefont {T.}~\bibnamefont {Fujisawa}},
  \bibinfo {author} {\bibfnamefont {S.}~\bibnamefont {Tarucha}}, \ and\
  \bibinfo {author} {\bibfnamefont {L.~P.}\ \bibnamefont {Kouwenhoven}},\
  }\href {\doibase 10.1103/RevModPhys.75.1} {\bibfield  {journal} {\bibinfo
  {journal} {Rev. Mod. Phys.}\ }\textbf {\bibinfo {volume} {75}},\ \bibinfo
  {pages} {1} (\bibinfo {year} {2002})}\BibitemShut {NoStop}%
\bibitem [{\citenamefont {Madelung}, \citenamefont {R\"ossler},\ and\
  \citenamefont {Schulz~(ed.)}(2002)}]{LandoltBornstein2002}%
  \BibitemOpen
  \bibfield  {author} {\bibinfo {author} {\bibfnamefont {O.}~\bibnamefont
  {Madelung}}, \bibinfo {author} {\bibfnamefont {U.}~\bibnamefont {R\"ossler}},
  \ and\ \bibinfo {author} {\bibfnamefont {M.}~\bibnamefont {Schulz~(ed.)}},\
  }\href@noop {} {\emph {\bibinfo {title} {Landolt-B\"ornstein - Group III
  Condensed Matter 41A1b}}}\ (\bibinfo  {publisher} {Springer-Verlag Berlin
  Heidelberg},\ \bibinfo {year} {2002})\BibitemShut {NoStop}%
\bibitem [{\citenamefont {Hanson}\ \emph {et~al.}(2007)\citenamefont {Hanson},
  \citenamefont {Kouwenhoven}, \citenamefont {Petta}, \citenamefont {Tarucha},\
  and\ \citenamefont {Vandersypen}}]{Hanson2007}%
  \BibitemOpen
  \bibfield  {author} {\bibinfo {author} {\bibfnamefont {R.}~\bibnamefont
  {Hanson}}, \bibinfo {author} {\bibfnamefont {L.~P.}\ \bibnamefont
  {Kouwenhoven}}, \bibinfo {author} {\bibfnamefont {J.~R.}\ \bibnamefont
  {Petta}}, \bibinfo {author} {\bibfnamefont {S.}~\bibnamefont {Tarucha}}, \
  and\ \bibinfo {author} {\bibfnamefont {L.~M.~K.}\ \bibnamefont
  {Vandersypen}},\ }\href@noop {} {\bibfield  {journal} {\bibinfo  {journal}
  {Rev. Mod. Phys.}\ }\textbf {\bibinfo {volume} {79}},\ \bibinfo {pages}
  {1217} (\bibinfo {year} {2007})}\BibitemShut {NoStop}%
\bibitem [{\citenamefont {Koppens}\ \emph {et~al.}(2005)\citenamefont
  {Koppens}, \citenamefont {Folk}, \citenamefont {Elzerman}, \citenamefont
  {Hanson}, \citenamefont {van Beveren}, \citenamefont {Vink}, \citenamefont
  {Tranitz}, \citenamefont {Wegscheider}, \citenamefont {Kouwenhoven},\ and\
  \citenamefont {Vandersypen}}]{Koppens2005}%
  \BibitemOpen
  \bibfield  {author} {\bibinfo {author} {\bibfnamefont {F.~H.~L.}\
  \bibnamefont {Koppens}}, \bibinfo {author} {\bibfnamefont {J.~A.}\
  \bibnamefont {Folk}}, \bibinfo {author} {\bibfnamefont {J.~M.}\ \bibnamefont
  {Elzerman}}, \bibinfo {author} {\bibfnamefont {R.}~\bibnamefont {Hanson}},
  \bibinfo {author} {\bibfnamefont {L.~H.~W.}\ \bibnamefont {van Beveren}},
  \bibinfo {author} {\bibfnamefont {I.~T.}\ \bibnamefont {Vink}}, \bibinfo
  {author} {\bibfnamefont {H.~P.}\ \bibnamefont {Tranitz}}, \bibinfo {author}
  {\bibfnamefont {W.}~\bibnamefont {Wegscheider}}, \bibinfo {author}
  {\bibfnamefont {L.~P.}\ \bibnamefont {Kouwenhoven}}, \ and\ \bibinfo {author}
  {\bibfnamefont {L.~M.~K.}\ \bibnamefont {Vandersypen}},\ }\href@noop {}
  {\bibfield  {journal} {\bibinfo  {journal} {Science}\ }\textbf {\bibinfo
  {volume} {309}},\ \bibinfo {pages} {1346} (\bibinfo {year}
  {2005})}\BibitemShut {NoStop}%
\bibitem [{\citenamefont {Barthel}\ \emph {et~al.}(2012)\citenamefont
  {Barthel}, \citenamefont {Medford}, \citenamefont {Bluhm}, \citenamefont
  {Yacoby}, \citenamefont {Marcus}, \citenamefont {Hanson},\ and\ \citenamefont
  {Gossard}}]{Barthel2012}%
  \BibitemOpen
  \bibfield  {author} {\bibinfo {author} {\bibfnamefont {C.}~\bibnamefont
  {Barthel}}, \bibinfo {author} {\bibfnamefont {J.}~\bibnamefont {Medford}},
  \bibinfo {author} {\bibfnamefont {H.}~\bibnamefont {Bluhm}}, \bibinfo
  {author} {\bibfnamefont {A.}~\bibnamefont {Yacoby}}, \bibinfo {author}
  {\bibfnamefont {C.~M.}\ \bibnamefont {Marcus}}, \bibinfo {author}
  {\bibfnamefont {M.~P.}\ \bibnamefont {Hanson}}, \ and\ \bibinfo {author}
  {\bibfnamefont {A.~C.}\ \bibnamefont {Gossard}},\ }\href {\doibase
  10.1103/PhysRevB.85.035306} {\bibfield  {journal} {\bibinfo  {journal} {Phys.
  Rev. B}\ }\textbf {\bibinfo {volume} {85}},\ \bibinfo {pages} {035306}
  (\bibinfo {year} {2012})}\BibitemShut {NoStop}%
\bibitem [{\citenamefont {Khaetskii}\ and\ \citenamefont
  {Nazarov}(2000)}]{Khaetskii2000}%
  \BibitemOpen
  \bibfield  {author} {\bibinfo {author} {\bibfnamefont {A.~V.}\ \bibnamefont
  {Khaetskii}}\ and\ \bibinfo {author} {\bibfnamefont {Y.~V.}\ \bibnamefont
  {Nazarov}},\ }\href {\doibase 10.1103/PhysRevB.61.12639} {\bibfield
  {journal} {\bibinfo  {journal} {Phys. Rev. B}\ }\textbf {\bibinfo {volume}
  {61}},\ \bibinfo {pages} {12639} (\bibinfo {year} {2000})}\BibitemShut
  {NoStop}%
\bibitem [{\citenamefont {Stepanenko}\ \emph {et~al.}(2012)\citenamefont
  {Stepanenko}, \citenamefont {Rudner}, \citenamefont {Halperin},\ and\
  \citenamefont {Loss}}]{Stepanenko2012}%
  \BibitemOpen
  \bibfield  {author} {\bibinfo {author} {\bibfnamefont {D.}~\bibnamefont
  {Stepanenko}}, \bibinfo {author} {\bibfnamefont {M.}~\bibnamefont {Rudner}},
  \bibinfo {author} {\bibfnamefont {B.~I.}\ \bibnamefont {Halperin}}, \ and\
  \bibinfo {author} {\bibfnamefont {D.}~\bibnamefont {Loss}},\ }\href@noop {}
  {\bibfield  {journal} {\bibinfo  {journal} {Phys. Rev. B}\ }\textbf {\bibinfo
  {volume} {85}},\ \bibinfo {pages} {075416} (\bibinfo {year}
  {2012})}\BibitemShut {NoStop}%
\bibitem [{\citenamefont {Danon}(2013)}]{Danon2013}%
  \BibitemOpen
  \bibfield  {author} {\bibinfo {author} {\bibfnamefont {J.}~\bibnamefont
  {Danon}},\ }\href {\doibase 10.1103/PhysRevB.88.075306} {\bibfield  {journal}
  {\bibinfo  {journal} {Phys. Rev. B}\ }\textbf {\bibinfo {volume} {88}},\
  \bibinfo {pages} {075306} (\bibinfo {year} {2013})}\BibitemShut {NoStop}%
\end{thebibliography}%

\end{document}